\documentclass[12pt]{article}
\usepackage{pdproc} 
\usepackage{epsfig}

\begin{document}

\renewcommand{\thefootnote}{\alph{footnote}}
  
\title{LENS AS A PROBE OF STERILE NEUTRINO MEDIATED OSCILLATIONS}

\author{C.~GRIEB, J.~M.~LINK, M.~L.~PITT, R.~S.~RAGHAVAN\footnote{Lecture delivered by R.~S.~Raghavan}, D.~ROUNTREE and R.~B.~VOGELAAR}

\address{Department of Physics and Institute of Particle, Nuclear and Astronomical Sciences \\
Virginia Polytechnic Institute and State University \\ 
Blacksburg, VA 24061, USA\\}

\abstract{
Sterile neutrino ($\nu_s$) conversion in meter scale baselines can be
sensitively probed using mono-energetic, sub-MeV, flavor pure $\nu_e$'s
from an artificial MCi source and the unique technology of the LENS low
energy solar $\nu_e$ detector. Active-sterile {\em oscillations} can be 
directly observed in the granular LENS detector itself to critically test 
and extend results of short baseline accelerator and reactor experiments.}
   
\normalsize\baselineskip=15pt

\section{Introduction}
Sterile neutrinos occur naturally in models of $\nu$ mass generation and figure in a variety of contemporary problems of particle physics, astrophysics and cosmology\cite{Kusenko:2006zc}.  Evidence for sterile $\nu$'s is thus of high interest, fueled by the LSND experiment that has claimed the appearance of $\bar{\nu}_e$ from a $\bar{\nu}_\mu$ beam at a baseline of 30~m.  The large $\Delta m^2$ ($\sim$1~eV$^2$) implied by LSND cannot be accommodated by the usual 3 active $\nu$'s with small mass splittings ~10$^{-5}$ to 10$^{-3}$~eV$^2$. An extended model of 3+1 mass eigenstates can explain the LSND effect as 
\begin{equation}
P_{\mu e} = \sin^2 2\theta_{\mu e}= 4U_{e4}^2U_{\mu 4}^2 
\end{equation}
where $U_{\alpha 4}$ are the mixing matrix elements with the fourth (mostly sterile) mass state in the 3+1 model.  If so, a complementary probe is $\bar{\nu}_e$ disappearance in reactor experiments, such as BUGEY\cite{Declais:1995su}, that have set limits on $\Delta m^2$ and $\sin^2 2\theta_{ee}= 4U_{e4}^2(1-U_{e4}^2)$.  These limits however, exclude most of the parameter space implied by LSND\cite{Sorel:2003hf}.  The two results do seem more compatible in a model of 3 active +2 sterile states\cite{Sorel:2003hf}.

The MiniBooNE-$\nu_{\mu}$ experiment\cite{Bazarko:1999hq} ($\nu_{\mu}\to\nu_e$) has now directly tested the LSND result, assuming CP invariance, and finds no evidence for LSND type oscillations\cite{Aguilar-Arevalo:2007it}.  Still combined fits find that it is yet possible to account for all the data in models with two or more sterile neutrinos and CP violation\cite{Maltoni:2007zf}.  In addition, non-standard physics, such as extra dimensions\cite{Pas:2006tk}, may account for all of the observed effects.  Finally, light sterile $\nu$'s, with mixing angles even smaller than LSND, can fix problems with r-process nucleosynthesis in supernovae\cite{McLaughlin:1999pd}.  In light of these scenarios it is desirable to push the search for large $\Delta m^2$ oscillations beyond LSND and MiniBooNE sensitivities.   

In the next step it is vital to {\em explicitly} observe active to sterile $\nu$ oscillations and measure the sterile $\nu$ mixing probabilities directly.  New tools are needed for these tasks. These objectives can be achieved by the unique technology of Low Energy Neutrino Spectroscopy (LENS) combined with a MCi, sub-MeV, mono-energetic, flavor-pure, $\nu_e$ source\cite{Grieb:2006mp}. 

\section{The LENS Detector}
\begin{figure}[tb]
\centerline{\epsfig{file=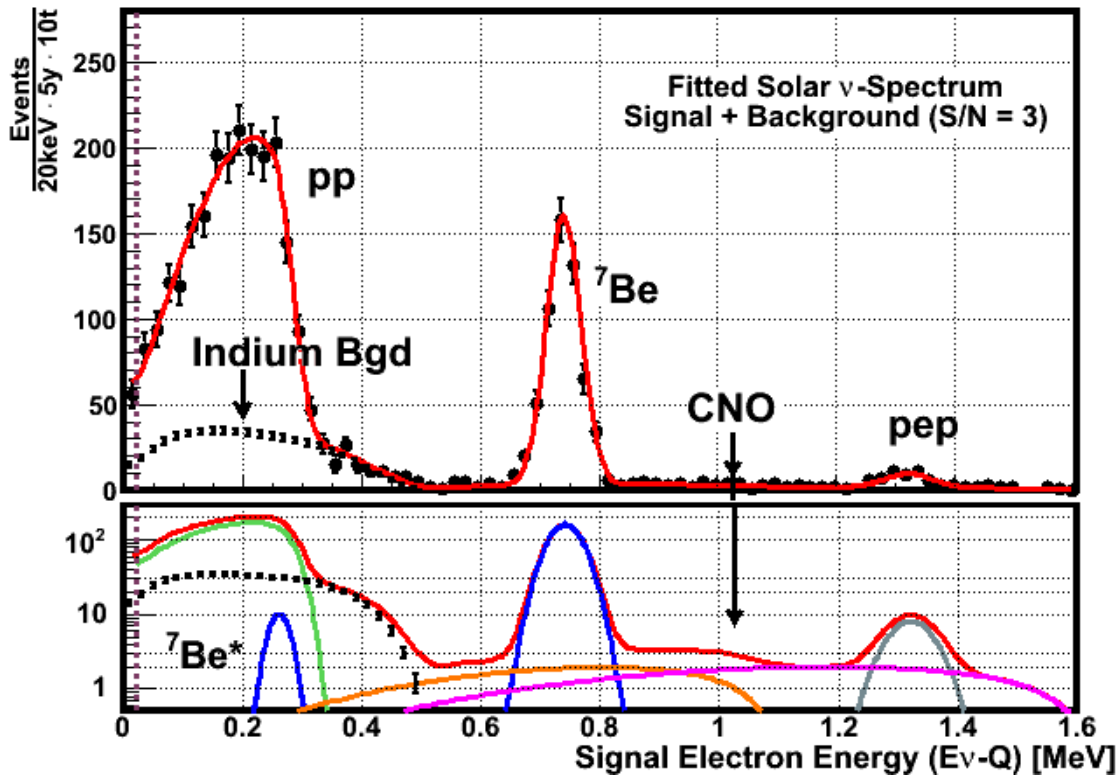,width=15.0cm}}
\caption{\label{solar_spect} Simulated 5 year-solar $\nu_e$ signal spectra in LENS\@. Energy spectrum (with $\sim$2000 pp events) at delays $<$10~$\mu$s, and random coincidences (from the pure $\beta$-decay of $^{115}$In target) at long delays.}
\end{figure}

The LENS detector is designed to observe low energy, electron neutrinos ($\nu_e$) and to measure their energy via a low threshold, tagged process. LENS $\nu_e$ detector is based on the charged current (CC) driven $\nu_e$ capture on $^{115}$In\cite{Raghavan:1976yc}:
\begin{eqnarray}
\nu_e  + ^{115}{\rm{In}}(95.6\%) & \rightarrow & ^{115}{\rm{Sn}}^* + e^- \\
& &\hspace{0.6cm}\raisebox{0.5em}{$\mid$}\!\!\!\to\ ^{115}{\rm{Sn}} + 2\gamma \nonumber
\end{eqnarray}
The $\nu_e$ capture leads to an isomeric state in $^{115}$Sn at 614 keV which emits a delayed ($\tau$ = 4.76~$\mu$s) 2$\gamma$ ($E = 116+498$~keV) cascade to the ground state of $^{115}$Sn. The reaction threshold is low with a $Q$-value of 114~keV\@. The $\nu_e$ signal energy, $E_e$ is directly related to the incident $\nu_e$ energy ($E_e= E_{\nu}-Q$).  The spectroscopic power of LENS is illustrated by its expected result on the low energy solar $\nu_e$ spectrum shown in Fig.~\ref{solar_spect}.

\begin{figure}[tb]
\centerline{\epsfig{file=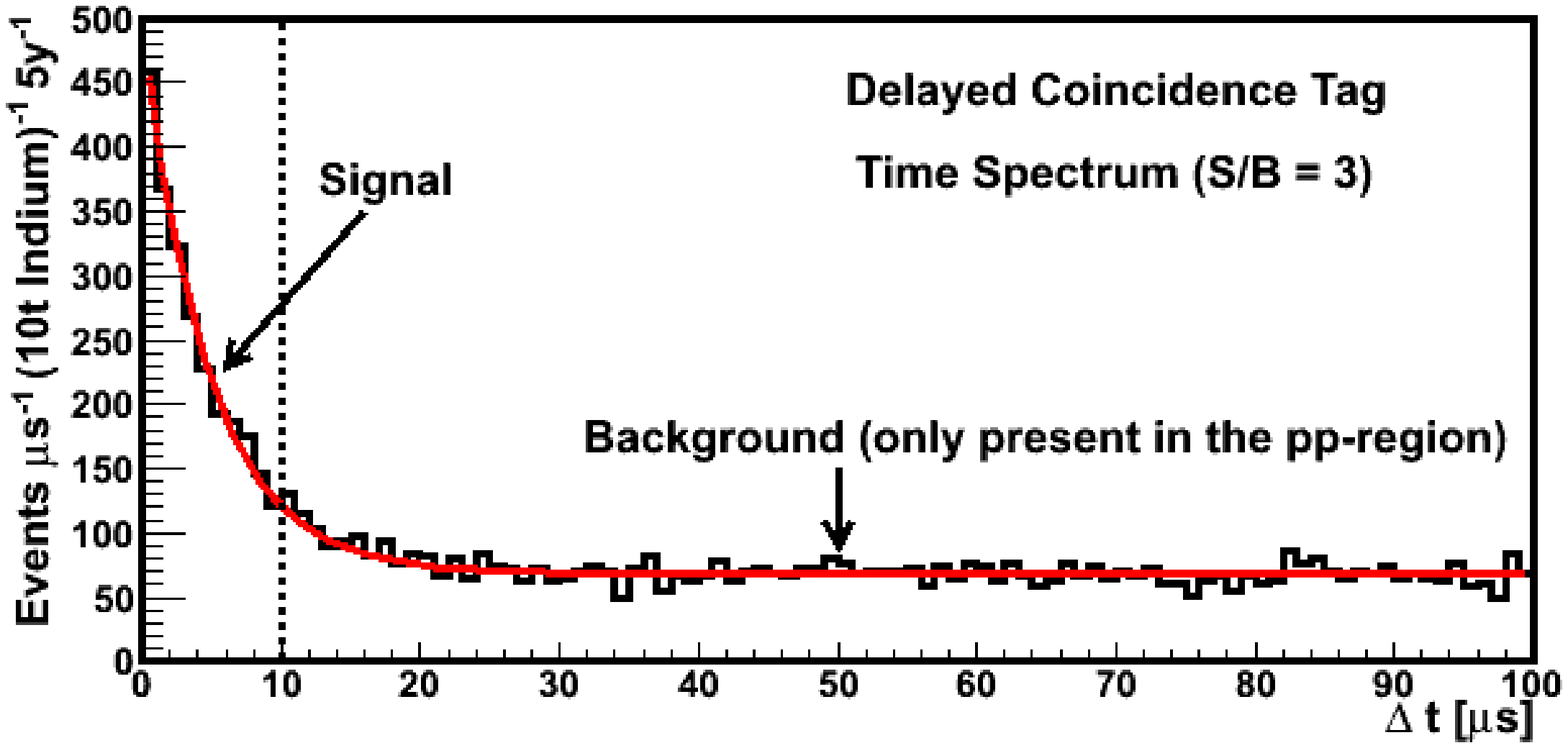,width=15.0cm}}
\caption{\label{time} Simulated 5 year-solar $\nu_e$ signal spectra in LENS\@. Delayed coincidence time spectrum fitted to the isomeric lifetime $\tau$=4.76~$\mu$s.}
\end{figure}

The delayed 2$\gamma$ signal is a powerful tag for the $\nu_e$ capture. The only valid events in LENS are {\em coincidence} events, the time distribution of which is shown in Fig.~\ref{time}. An exponential decay fit (with the signature lifetime of $^{115}$Sn$^*$) to the time spectrum leads to the true signal events that occur at early delays $\sim$10~$\mu$s and the background from random coincidences which are flat in decay time. This background arises dominantly via random coincidences with $\beta$'s from the natural decay of the target $^{115}$In with the end point at 498(4)~keV\@. It presents a severe (but soluble) problem for solar pp $\nu_e$ ($E_e<$300~keV, see Fig.~\ref{solar_spect}) but it is far less problematic for $E_e>$500~keV (e.g. for $^7$Be solar neutrinos). The background is, in any case, {\em measured} at long delays separately and concurrently. The $\nu_e$ event analysis tests the validity of a candidate tag by the time and space coincidence as well as the detailed template of the 2$\gamma$ tag shower. This determines the coincidence efficiency, modeled to be $>$85\% for primary signals $>$500~keV.

The detection medium in LENS is a liquid scintillator chemically loaded with indium (InLS). This technology uses a robust, well tested (Bell Labs\cite{Raghavan:2004}, BNL, LNGS and Virginia Tech) procedure that produces high quality InLS (In loading of 8-10\% by weight, with a scintillation output of $\sim$8000~h$\nu$/MeV and signal light attenuation length L(1/e) of 8~m that has been stable for $>$1 year)\cite{lens:2006}. InLS with up to 15\% In has been produced with promising properties and is currently in development. Anticipating positive results, we consider both 8\% and 15\% In loading.

The granular LENS detector is based on a novel ``scintillation lattice chamber" design\cite{lens:2006}. In this design the detector volume is optically segmented into cubic cells ($7.5\!\times\!7.5\!\times\!7.5$~cm) by a 3-dimensional cage of thin ($\sim$0.1~mm), double layered, transparent foils which contain a material with a smaller index of refraction than the scintillator (Table~\ref{mat_indx} lists materials that have been considered).  The scintillation light is then channeled primarily along the 3 coordinate axes centered on the vertex cell of the event. Coincident hits on phototubes at the end of the channels {\em digitally} determine the 3-D location of the event to a precision set by the cell size. Each cell is always viewed by the same combination of 6 or less phototubes. Thus LENS is, in effect, a 3-D array of $10^5$ nuclear counters with bench top sensitivity. 

\begin{table}[b!]
\begin{center}
\begin{tabular}{|c|c|c|c|} \hline
Material & Index of Refraction & Critical Angle & Phase \\
\hline
Ideal          & 1.15 & 45.0$^o$ & ---    \\ 
Air            & 1.00 & 37.8$^o$ & gas    \\ 
Teflon         & 1.35 & 55.9$^o$ & solid  \\
Water          & 1.33 & 54.7$^o$ & liquid \\
perflurohexane & 1.27 & 51.2$^o$ & liquid \\ \hline
\end{tabular}
\caption{\label{mat_indx} Materials that have been considered for light channeling through total internal reflection.  Critical angles greater than 45$^o$ result in light leakage, and angles less than 45$^o$ result in trapped light.}
\end{center}
\end{table}

Such a detector is ideally suited for the detection and energy spectrum measurement of solar neutrinos down to well below the pp endpoint.  In fact the LENS technology is the only potentially viable detector yet proposed for this purpose.  

The anticipated solar neutrino spectrum for a five year run is shown in Fig.~\ref{solar_spect}.  This spectrum assumes no neutrino oscillations.  Of course we know that neutrinos do oscillate, thus with the energy resolution available with LENS details of this oscillation physics including the interaction with matter can be determined.  In addition, LENS will open a new window on dynamics in the solar core.  For example, the pp neutrino spectrum may be used to derive the temperature profile of the sun\cite{Grieb:2006zc}.

\section{Sterile Oscillations}
The unique low energy threshold, energy resolution and spatial resolution of LENS provide an interesting opportunity to study neutrino oscillations from terrestrial sources, especially if one considers high $\Delta m^2$ oscillations, mediated by sterile neutrinos, as implied by the LSND experiment\cite{Athanassopoulos:1995iw}.

The LENS technology thus extends the power of tagged $\nu$ detection from the case of broadband $\bar{\nu}_e$ beam with $E_{\nu} > 1.8$~MeV in reactor experiments to mono-energetic {\em sub-MeV} $\nu_e$ that lead to explicit, meter scale oscillations directly detectable in modest sized granular detectors. Most issues in shape analysis of $P_{ee}$ and systematic normalization errors endemic to reactor $\nu$ experiments are thus avoided in LENS\@. A $\nu_e$ test allows direct comparison to MiniBooNE ($\nu_{\mu}$) without invoking CP invariance.

With the source placed at the center of detector, spherical $\nu_e$ oscillation wave can be observed by the radial variation $P_{ee}(R)$ through the disappearance of $\nu_e$. The LENS experiment thus uses $10^5$ detectors to trace the $\nu_e$ oscillation profile in detail. The result is thus far more transparent than that from usual methods with a few static or movable detectors.  As in the reactor experiment theoretical uncertainties due to $\nu$ cross section are common to all ``detectors" and thus cancel.  In addition the uncertainty from indium concentration cancels, or at least should average out because the scintillator volume is fully connected, and detection efficiency can be made flat across the detector because the mono-energetic signal allows triggering thresholds to be set to tune out any instability.

Several MCi sources of $^{51}$Cr have been produced and used to calibrate the Ga radiochemical solar $\nu_e$ detectors (1.7~MCi in GALLEX\cite{Cribier:1996cq} and 0.52~MCi in SAGE~\cite{Abdurashitov:1996dp}). Studies show that the production of $\sim$10~MCi $^{51}$Cr sources is feasible. The electron-capture decay ($\tau$ = 40~d) of $^{51}$Cr emits a mono-energetic $\nu_e$ of energy 0.753~MeV (90\%) which produces a $\nu_e$ signal at 0.639~MeV in LENS\@. We note that a MCi source experiment for precision calibration of the $^{115}$In $\nu_e$ capture cross-section is already part of the LENS program as LENS-CAL\cite{lens-cal:2006}.  In addition to $^{51}$Cr other source across a wide range of energies are possible.  These source are listed in table~\ref{sources}.

\begin{table}[b!]
\begin{center}
\begin{tabular}{|c|c|c|} \hline
Source & $\nu$ Energy & Half-life \\
\hline
$^{51}$Cr & 751 keV  & 27.7 d \\ 
$^{37}$Ar & 814 keV  & 34.9 d \\ 
$^{65}$Zn & 1.35 MeV & 244 d  \\
$^{55}$Fe & 232 keV  & 2.34 y \\ \hline
\end{tabular}
\caption{\label{sources} Possible electron caputre $\nu_e$ sources.}
\end{center}
\end{table}

The source experiment requires a $\sim 4\pi$ source-detector geometry for maximum sensitivity. The primary design criterion is then the background arising from hard $\gamma$'s of impurity activities in the source (the only $\gamma$ from $^{51}$Cr itself is of energy 0.32~MeV, easily shielded and far below the $\nu_e$ signal energy). The source will be encased in a heavy-metal shielding container to cut down radiation outside the container below permissible safety limits. The SAGE data\cite{Abdurashitov:1996dp} show that the dominant $\gamma$-rays outside the container are from the impurity activity $^{46}$Sc (3~Ci/MCi $^{51}$Cr) of 2 hard $\gamma$'s (1.12 and 0.889~MeV). In a $\sim$25~cm radius tungsten container this releases $\sim10^3$~$\gamma$/s into the detector, $\ll 10^6$/s, the benchmark rate of $^{115}$In $\beta$'s that underlies all random coincidence background considerations in LENS\cite{lens:2006}. Thus, a 10~MCi $^{51}$Cr source in a $\sim$50~cm diameter heavy-metal sphere at the center of the LENS detector is a viable concept with state-of-the art technology.

\begin{figure}[tb]
\centerline{\epsfig{file=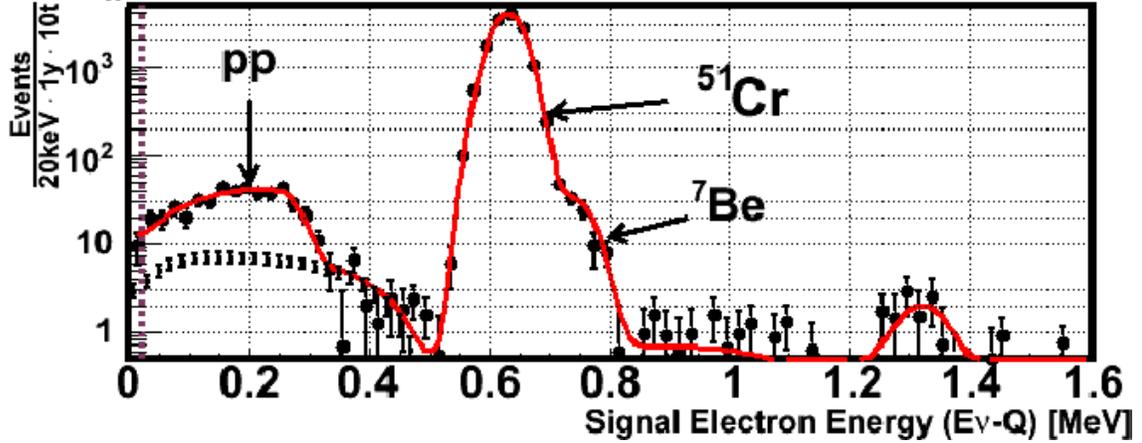,width=15.0cm}}
\caption{\label{source} Spectrum from 4$\times$100~day exposure to a 10~MCi source of $^{51}$Cr with the scaled solar spectrum included. The Cr signal has $1.3\times10^4$ events with $\sim$0.1\% background (from the $^7$Be line).}
\end{figure}

The $\nu_e$ spectroscopy of $^{51}$Cr ($4\times100$~day exposure to a 10~MCi source) in a full scale LENS detector is shown in the bottom panel of Fig.~\ref{source} (notice the log scale).  The Cr signal ($\sim$13,300 events) is $\sim$1000 times the solar ``background".  An internal MCi source in a real-time rare event counting detector is, by itself, a major first (SAGE/GALLEX are
radio-chemical activation experiments).

\begin{table}[t!]
\begin{center}
\begin{tabular}{|c|c|c|c|} \hline
Model & $\Delta m^2_{ij}$ (eV$^2$) & $U^2$ & $\sin^2 2\theta_{ee}$ \\
 & & & $=\!4U^2(1\!-\!U^2)$ \\
\hline
3+1  & 0.92$_{14}$ & 0.0185 & 0.073     \\ \hline
3+2a & 0.92$_{14}$ & 0.0146 & 0.057     \\ 
     & 22.1$_{15}$ & 0.0013 & 0.005     \\ \hline
3+2b & 0.46$_{14}$ & 0.0081 & 0.032     \\
     & 0.89$_{15}$ & 0.0156 & 0.062     \\ \hline
\end{tabular}
\caption{\label{tab1} Active-Sterile splittings and mixing parameters compatible with LSND and the null short baseline data\protect\cite{Sorel:2003hf}, as used in the analysis shown in Fig.~\ref{dmsq_source}.}
\end{center}
\end{table}

\begin{figure*}[t!]
\centerline{\epsfig{file=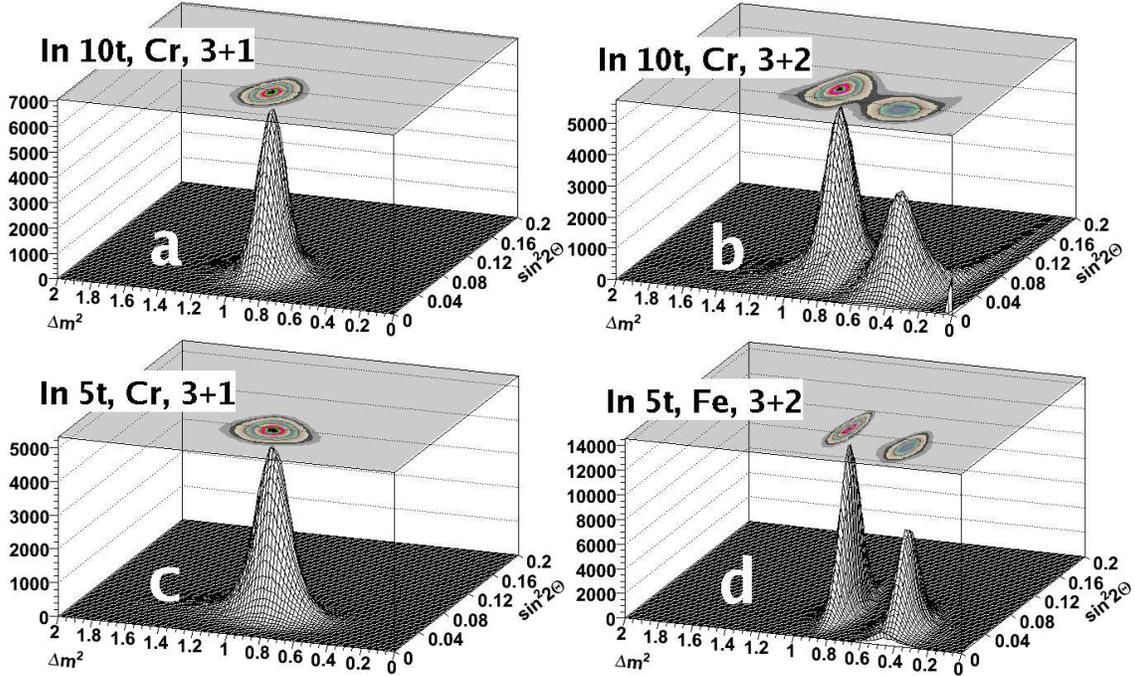,width=15cm}}
\caption{\label{dmsq_source} Analysis of 3+1 and 3+2 models of sterile neutrinos in LENS\@.  In 10t=8\% In LENS-Sol ({\bf A}); In 5t=15\% In LENS-Sterile ({\bf B}); Cr=4~MCi$\times$100~days exposure to $^{51}$Cr; and Fe=10~MCi$\times$2~years exposure to $^{55}$Fe.}
\end{figure*}

In 3+2 models, with mixing angles and oscillation frequencies in flavor states $\alpha=e,\ \mu,\ \tau, s_{1,2}$, mass states n=1-5 where states 4 and 5 are mostly sterile and $U_{\alpha n}$ are the elements of the $\nu$ mixing matrix, the e-flavor survival, $P_{ee}$, is:
\begin{equation}
P_{ee} \simeq 1\!-\!4U_{e4}^2(1\!-\!U_{e4}^2)\sin^2 x_{41}\!-\!
                    4U_{e5}^2(1\!-\!U_{e5}^2)\sin^2 x_{51}
\end{equation}
where cross terms such as $U_{e4}^2U_{e5}^2$ are neglected, and $x_{ij} = 
1.27\Delta m_{ij}^2{\rm(eV^2)} L{\rm(m)}/E_{\nu}{\rm(MeV)}$. The values of $U_{ei}$ and $\Delta m^2$ (shown in Table~\ref{tab1}) are from Ref\cite{Sorel:2003hf}. With $\Delta m^2 = 1$~eV and $E_{\nu}\sim 0.753$~MeV (from $^{51}$Cr), full flavor recovery occurs in $\sim$2~m, directly observable in a lab-scale detector.

We consider two detector configurations, {\bf A} and {\bf B} (Table~\ref{tab2}), based on the lattice array design with $7.5\!\times\!7.5\!\times\!7.5$~cm cells. Design {\bf A} is the full scale solar detector (LENS-Sol 8\% In). The smaller detector {\bf B}, anticipates further development of a higher density (15\% In) InLS to exploit the linear dependence of the event rate with In density. It is thus specific for sterile $\nu$ search and costs $\sim$50\% less than {\bf A}\@. With comparable (density $\times$ In mass), the event rate is comparable for a source assembly at the center of {\bf A} or {\bf B} in a 50~cm spherical volume.

\begin{table}[b!]
\begin{center}
\begin{tabular}{|l|c|c|c|c|} \hline
Configuration & $\rho_{In}$ & $d_{detector}$ & $m_{In}$ & $m_{total}$ \\
 & (wt. \%)    & (meters)       & (tons)   & (tons)      \\ \hline
 {\bf A} -- LENS-Sol     & 8  & 5.1 & 9.9 & 125 \\ \hline
 {\bf B} -- LENS-Sterile & 15 & 3.3 & 5.1 & 34  \\ \hline
\end{tabular}
\caption{\label{tab2} Design options for LENS sterile $\nu$ search.}
\end{center}
\end{table}

The sensitivity of radial distributions of events in designs {\bf A} and {\bf B} to sterile $\nu$ oscillation parameters was analyzed by a Monte Carlo technique. The number of $\nu_e$ detected, $N_{0i}$ is calculated for each cubic cell element $i$ of the detector assuming no oscillations. Then, assuming an oscillation model of (3+1) or (3+2) with parameters ($\Delta m^2$, $\sin^2 2\theta$) or ($\Delta m_1^2$, $\sin^2 2\theta_1$; $\Delta m_2^2$, $\sin^2 2\theta_2$) the number of $\nu_e$ detected in each cell, $N_i$, is calculated. The cells are then grouped into $k$ bins according to their distance, $d_i$, from the source. A random sample data set for one experiment is created under the oscillation assumption. Then the theoretical event ratio as a function of $d$, $N_i / N_{0i} = f(\sin^2 2\theta, \Delta m^2, d)$, is fitted to the sample data set with the oscillation parameters as free parameters. The process is repeated $10^6$ times with random data sets and each resulting set of fitted parameters $(\sin^2 2\theta, \Delta m^2)_i$ is stored.  The distribution of these values (see Fig.~\ref{dmsq_source}) is a direct measure of the statistical uncertainty of ($\sin^2 2\theta$, $\Delta m^2$) in a single measurement.

\begin{figure}[t]
\centerline{\epsfig{file=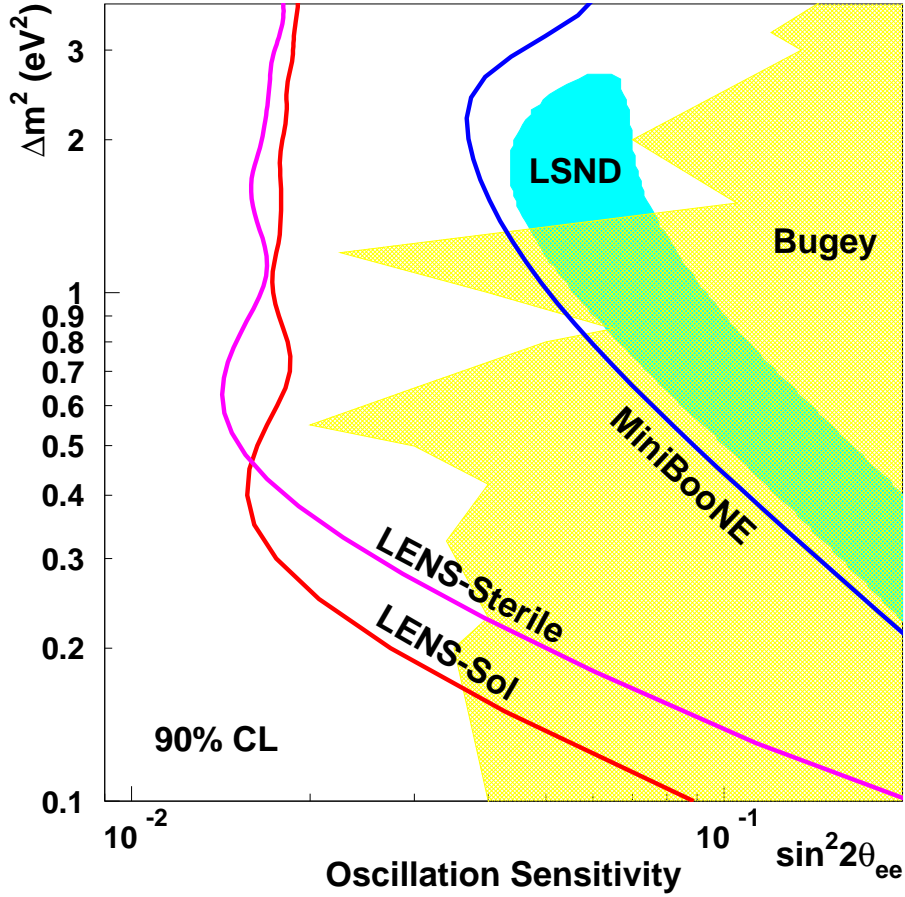,width=12.0cm}}
\caption{\label{fig3} Exclusion plots of sensitivity to active-sterile oscillations in LENS (with 4$\times$100~day exposure to 10~MCi Cr source) compared to the allowed region for LSND and the limits from BUGEY\protect\cite{Declais:1995su}, and MiniBooNE\protect\cite{Aguilar-Arevalo:2007it}\@.  LSND and MiniBooNE are plotted assuming the best fit values for $U_{e4}$ and $U_{\mu 4}$ from Sorel {\em et al.}\protect\cite{Sorel:2003hf}.}
\end{figure}

The models in Table~\ref{tab1} predict 2 main effects: 1) the minimum oscillation amplitude of $\sim$5\% occurs with a single frequency for 3+1 and (practically also in 3+2a); 2) two frequencies occur in 3+2b. In the analysis, we test designs {\bf A} and {\bf B} for 1) a single frequency and 2) double frequency as in 3+2b. Figs.~\ref{dmsq_source}a,b for design {\bf A} (full scale solar detector) show that a 5\% single frequency oscillation (3+1, 3+2a) as well as the double frequency oscillation for 3+2b can be clearly observed with $4\times100$~day exposure to a 10~MCi $^{51}$Cr source. The mixing parameter $\sin^2 2\theta$ can be determined with $1\sigma$ precision of 25\% in both cases. In the smaller detector {\bf B} the lower of the double frequency is not picked up with Cr because the radial dimension of {\bf B} is insufficient to catch the $P_{ee}$ recovery. In this case, another source, $^{55}$Fe with $E_{\nu} = 236$~keV, can be used. A ~2-year exposure to a 10~MCi Fe source ($\tau$ = 3.8~y) (Fig.~\ref{dmsq_source}d), shows excellent resolution of the double frequency effect. A MCi Fe source has not been produced so far, thus, this technology awaits {\em ab initio} development.

\begin{figure}[b!]
\centerline{\epsfig{file=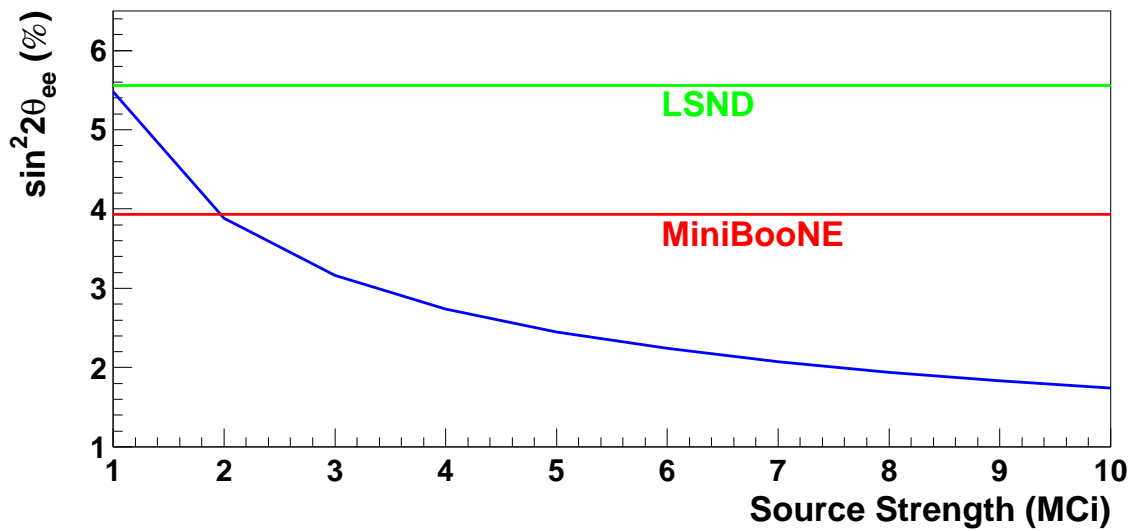,width=15.0cm}}
\caption{\label{strength} Sensitivity to sterile oscillations vs $^{51}$Cr source strength.  The MiniBooNE limit at 1~eV$^2$ can be achieved with a source of less than 2~MCi.}
\end{figure}

Fig.~\ref{fig3} shows the exclusion plots of $\Delta m^2$ vs. $\sin^2 2\theta_{ee}$ for 90\% confidence level sensitivity to oscillations in the designs {\bf A} and {\bf B} compared to those from BUGEY, LSND and MiniBooNE.  Fig.~\ref{strength} shows that, indeed, a relatively modest 3~MCi source of $^{51}$Cr is sufficient to reach the projected oscillation sensitivity of MiniBoonNE.

These analyses show that the LENS approach in {\bf A} or {\bf B} can exclude parameter regions of active-sterile oscillations significantly beyond those of BUGEY, LSND and MiniBooNE. Further, the LENS approach is complementary to the proposed NC-based $\nu_{\mu}$ disappearance test which is sensitive to $\sin^2 2\theta_{\mu\mu}$\cite{Garvey:2005pn}.

In summary, the LENS technology offers a novel attack for discovery of active-sterile oscillations with sensitivities well beyond those in LSND, MiniBooNE and BUGEY\@. Design {\bf B} offers the opportunity to accomplish these goals in a relatively modest detector that can map the full oscillation as a function of radius. However, R\&D on the high density InLS needs to be completed and the technology of the Fe source developed {\em ab initio}. The straightforward approach would use the full scale LENS detector, design A, for which the Cr source and detector technologies are already well developed. The next step in the latter is the construction and operational tests of the prototype detector MINILENS which is now being initiated.

\bibliography{lens_sbno}
\bibliographystyle{test}

\end{document}